\documentclass[12pt]{article}

\usepackage{epsfig}

\usepackage{amssymb}
\usepackage{amsfonts}

\usepackage{color}
\def\be{\begin{equation}}
\def\ee{\end{equation}}
\def\ba{\begin{array}{c}}
\def\ea{\end{array}}
\def\p{\partial}
\def\ben{$$}
\def\een{$$}

\newcommand{\bea}{\begin{eqnarray}}
\newcommand{\eea}{\end{eqnarray}}

\newcommand{\bbr}{\br\!\br}
\newcommand{\kkt}{\kt\!\kt}

\newcommand{\kt}{\rangle}
\newcommand{\br}{\langle}

\newtheorem{thm}{Theorem}
\newtheorem{cor}[thm]{Corollary}
\newtheorem{lemma}[thm]{Lemma}

\begin{document}

\begin{center}

{\Large \bf

Supersymmetry and exceptional points

}

\vspace{0.8cm}

  {\bf Miloslav Znojil}

\vspace{0.2cm}

The Czech Academy of Sciences, Nuclear Physics Institute,

 Hlavn\'{\i} 130,
250 68 \v{R}e\v{z}, Czech Republic

\vspace{0.2cm}

 and

\vspace{0.2cm}

Department of Physics, Faculty of Science, University of Hradec
Kr\'{a}lov\'{e},

Rokitansk\'{e}ho 62, 50003 Hradec Kr\'{a}lov\'{e},
 Czech Republic

\vspace{0.2cm}

{e-mail: znojil@ujf.cas.cz}

%
%
%\vspace{1mm} Nuclear Physics Institute of the CAS, Hlavn\'{\i} 130,
%250 68 \v{R}e\v{z}, Czech Republic
%
%

\end{center}

%\newpage

\subsection*{Keywords}
.

hiddenly Hermitian quantum Hamiltonians;

Kato exceptional points;

regularized singular potentials;

hiddenly Hermitian supersymmetries;

infinite family of regularized harmonic oscillators

%\newpage

\section*{Abstract}

The phenomenon of degeneracy of energy levels
is often attributed either to an underlying (super)symmetry
(SUSY), or to the presence of a Kato exceptional point (EP).
In our paper a conceptual bridge between the two notions is
proposed to be provided by
the recent upgrade
of the basic principles of quantum theory
called, equivalently,
${\cal PT}-$symmetric
or three-Hilbert-space (3HS) or quasi-Hermitian
formulation in the current physical literature.
Although
the original purpose of the 3HS approach
laid in the mere simplification of technicalities,
it is shown here to serve also as a natural
theoretical link between the
apparently remote concepts of EPs and SUSY.
An explicit illustration of their close mutual interplay is
provided by the description of
infinitely many supersymmetric,
mutually non-equivalent and EP-separated
regularized spiked harmonic oscillators.

\newpage

\section{Introduction}

Ten years ago people perceived the concepts of
supersymmetry and of exceptional points as the two remote if not
fully separate
subjects of research. For example,
during
the international
workshop ``Supersymmetric Quantum Mechanics and Spectral Design''
(July 18 - 30, 2010, Benasque, Spain, organized by
A. Andrianov et al \cite{benasque},
with proceedings \cite{benasqueproc}),
just one talk was devoted to the role of exceptional points
in supersymmetric quantum systems.
In parallel,
no speaker mentioned supersymmetry
during the workshop
``Exceptional points in physics'' (November 2 - 5, 2010,
NITheP, Stellenbosch, South Africa,
organized by W. D. Heiss \cite{Stellenbosch}).
Also the same year,
the participants of the 9th continuation of the series
``Pseudo-Hermitian Hamiltonians
in Quantum Physics'' (June 21 - 24, 2010, Zhejiang University,
Hangzhou, China, organized by
J.-D. Wu \cite{junde}, with proceedings \cite{jundeproc}) did not yet
identify their field of interest as an
emergent bridge between the two concepts, from either the
mathematical and physical points of view.

At present, the situation is different.
The use of the
mathematical notion of exceptional points
(EPs, \cite{Kato})
makes an impression of being, in multiple branches of
physics,  ubiquitous \cite{Stellenboschb,Christodoulides}.
The same growth of popularity became also characteristic
for the concept of  hidden Hermiticity {\it alias\,}
pseudo-Hermiticity \cite{ali}
or, better,
quasi-Hermiticity \cite{Dieudonne,Geyer},
most widely known under the physics-emphasizing
nicknames of ``${\cal PT}-$symmetry'' \cite{Carl}
or ``gain-loss balance'' \cite{Carlbook}.
In comparison, the concept of supersymmetry (SUSY, \cite{Duplij})
seems, undeservedly,
overshadowed
by these developments.

In our present paper we intend to demonstrate that
the concept of SUSY witnesses a revival, in particular,
due to its remarkable overlap with both of the
EP- and ${\cal PT}-$related areas of research.
The new conceptual bridges are emerging between these notions
in a way which we intend to describe in what follows.
On the SUSY side,
in essence, the core of the mutual influence lies in the
emergence of
the new, unconventional
realizations of the SUSY algebra in quantum mechanics.
One witnesses there a shift of attention
from Hermitian to non-Hermitian Hamiltonians.
On the EP side, in parallel, precisely such a shift makes the
EP singularities accessible.
Indeed, they are suddenly found as lying on boundaries of the
relevant parametric domains admitting the standard probabilistic
interpretation of the systems.

Our detailed scrutiny of this territory will proceed
step by step.
In section~\ref{section2}
an introductory review
of the formalism
will be provided
summarizing
the basic features
of the  ${\cal PT}-$emphasizing quantum mechanics of closed systems.
Subsequently,
in section~\ref{section3}  we will turn attention
to a few most relevant
features of the role of EPs in such a theory or theories.
In section \ref{section4} we will finally
combine the preceding material
in a climax outlining the bridges between SUSY and EPs.
We will show, in essence,  that the
recent transfer of the applications of
SUSY to quasi-Hermitian quantum systems
extends the class of eligible models while still
finding its natural limitations of applicability precisely
at the same natural EP boundaries.

A few complementary aspects of our message will be also
discussed in section~\ref{section5}.
We will emphasize there that there exists
a close relationship between the
EP-related properties of the SUSY and non-SUSY quantum systems.
All of these systems
share the limitations of their
unitarity and observability (i.e.,
of their probabilistic interpretation)
strictly along the mathematical
parameter-range
boundaries formed by the EP
singularities.

%\newpage

\section{The concept of hiddenly Hermitian Hamiltonians\label{section2}}

In the present paper our
perception of quantum physics will be restricted
to the theory in which a
non-Hermitian but hiddenly Hermitian Hamiltonian $H$ generates
a {\em strictly unitary\,} evolution of quantum system
in  question
(see
the resolution of an apparent paradox as given in
the early review of this approach to quantum mechanics
in \cite{Geyer}).

\subsection{Non-Hermitian physics: closed {\it versus} open systems}

In a concise introductory comment let us
point out that
the quantum systems under our present consideration are
usually called ``closed systems''.
Carefully,
one has to separate the study of these models (with real energies)
from the descriptions of
other, manifestly non-unitary quantum systems (with complex energies)
called ``open''.
Unfortunately, very similar language
is often used
in the literature on both the closed and
open quantum dynamics.
In a word of warning against possible misunderstandings let us
recall, for illustration, the
existence of potentially misleading
overlaps in terminology.
Often,
people indiscriminately speak about
``non-Hermitian quantum mechanics''
having in mind {\em either\,} the resonances in unstable,
open quantum systems \cite{Nimrod,Ingrid}
{\em or\,} the
safely stable
bound states in a closed, unitary setup \cite{Carl}.

Needless to add that the
specific trademark logo of
${\cal PT}-$symmetry
born in the context of quantum
mechanics \cite{BG,BB} and quantum field theory
\cite{BM} became quickly popular also
in the non-quantum world \cite{Carlbook}.
In optics, for example, the idea
proved particularly
successful \cite{cartoon}. Many research teams using
Maxwell equations revealed that the propagation of light
acquires remarkable properties when studied in a
non-conservative medium
characterized by
${\cal PT}-$symmetry redefined, {\it ad hoc}, as
a balance between loss and gain.
Thirdly, the same name ``${\cal PT}-$symmetry''
extended its implementations also to non-linear systems
and equations,
with the deepest relevance and
applicability even in nanotechnologies
\cite{Christodoulides}.

Out of such a broad area of physics,
our present attention will only be paid to the
quantum closed systems.

\subsection{Triplets of Hilbert spaces: A key to the unitarity paradox}

The famous Stone's theorem \cite{Stone} implies that
in a preselected Hilbert space (denoted, say,
by a dedicated symbol ${\cal K}$), the
Schr\"{o}dinger equation
 \be
 {\rm i}\frac{d}{dt}\,|\psi(t)\kt = H\,|\psi(t)\kt
 \label{coxa}
 \ee
cannot describe a unitary (i.e., norm-preserving)
evolution of a state
$|\psi(t)\kt$ whenever the Hamiltonian
happens to be non-Hermitian, $H \neq H^\dagger$.
Fortunately, in a way dating back to Dyson's studies
of ferromagnetism \cite{Dyson} it became clear that
in many realistic quantum models
the non-unitarity
of the evolution in  ${\cal K}$
may be declared, under certain conditions, irrelevant.
Indeed, in many practical applications,
it may make sense to treat $H$ as a mere auxiliary
isospectral avatar
of a ``true and correct'' Hamiltonian
 \be
 \mathfrak{h}=\Omega\,H\,\Omega^{-1}
 \label{precoxa}
 \ee
which is ``acceptable''
(i.e., self-adjoint, $\mathfrak{h}=\mathfrak{h}^\dagger$)
in {\em another},
true and correct, {\em physical} Hilbert space of states
(to be denoted here
by another dedicated symbol ${\cal L}$).
In other words, the Dyson-attributed  preconditioning~(\ref{precoxa})
changes the picture while making the theory
compatible with the Stone's theorem as well as with the
conventional-textbook closed-system unitary-evolution requirements.

Dyson's numerous followers (cf., e.g., review \cite{Geyer})
were pragmatic practitioners who
usually started
their considerations from the principle of correspondence.
Having specified a realistic
(typically, a fermionic atomic-nucleus-model)
Hamiltonian $\mathfrak{h}$ acting in ${\cal L}$
they needed to evaluate its low-lying spectrum.
In such a (typically, variational) setting
they revealed that the calculations may be perceivably simplified,
via preconditioning (\ref{precoxa}),
after an educated {\em ad hoc\,}
guess of a suitable invertible Hermiticity-breaking
map $\Omega^{-1}: {\cal L} \to {\cal K}$.
This map represented, typically,
correlations between fermions living in ${\cal L}$
in the language of certain
effective bosons living in ${\cal K}$.
In practice, such a strategy really accelerated the calculations
yielding
a perceivably simplified picture
of the underlying physical reality.

The Dyson-recommended
preconditioning $\mathfrak{h}\to H$
is, almost without exceptions, motivated by the
simplification of the solution of
Schr\"{o}dinger Eq.~(\ref{coxa}).
Naturally, the price to pay is not entirely
negligible because the non-Hermiticity of $H$
(which must still be assumed diagonalizable \cite{ali})
implies that its right and left eigenvectors
are not just mutual Hermitian conjugates.
For this reason (see also \cite{SIGMA}) it makes sense
to complement the
bra-vector conjugates $\br n|$ of the
conventional
right eigenkets
$|n\kt$
by the
left-eigenbras of $H$
(denoted as eigenbrabras $\bbr n|$)
with, naturally, {\em different\,} conjugates $|n\kkt \neq |n\kt$
(that's why we amended the notation).

For the same reason the
full and explicit description of a pure
time-dependent state
of the systems
requires not only the knowledge of
the ket-state solutions
$|\psi(t)\kt$ of the evolution equation (\ref{coxa})
but also the complementary,
non-Hermiticity-related  knowledge
of the brabra (or ketket) solutions
of the following complementary evolution
equation in ${\cal K}$,
 \be
 {\rm i}\frac{d}{dt}\,|\psi(t)\kkt = H^\dagger\,|\psi(t)\kkt\,.
 \label{doxa}
 \ee
The two Schr\"{o}dinger equations
are interrelated by Eq.~(\ref{precoxa})
and
by the fact that  $\mathfrak{h}=\mathfrak{h}^\dagger$
in ${\cal L}$. Thus, in ${\cal K}$ we can
speak about quasi-Hermiticty of $H$ \cite{Dieudonne}
and write
 \be
 H^\dagger \Theta=\Theta\,H\,,\ \ \ \  \ \ \
 \Theta=\Omega^\dagger\Omega\,.
 \label{dieux}
 \ee
In the stationary cases
this observation makes the second
Schr\"{o}dinger
equation redundant because we may set, simply,
$|\psi(t)\kkt=\Theta\,|\psi(t)\kt$.
Moreover, after we introduce such
a biorthonormal basis $\{|n\kkt,|n\kt\}$ in ${\cal K}$
which diagonalizes $H$, we have relations
 \be
 H=\sum_n\,|n\kt\,E_n\bbr n|\,,
 \ \ \ \
 I= \sum_n\,|n\kt\,\bbr n|\,.
 \label{bio0}
 \ee
We become able to define the third Hilbert space ${\cal H}$
when changing the inner product $(\bullet,\bullet)$ between
any two elements $\psi_a$ and $\psi_b$ in ${\cal K}$.
Thus, once we define
 \be
 (\psi_a,\psi_b)_{\cal K}=\br \psi_a|\psi_b \kt\,,
 \ \ \ \ \
 (\psi_a,\psi_b)_{\cal H}=\br \psi_a|\Theta|\psi_b \kt\,,
 \ \ \ \
 \Theta= \sum_n\,|n\kkt\,\bbr n|\,
 \label{innpro}
 \ee
we may conclude that
there are no reasons for calling
$H$ ``non-Hermitian''. It is manifestly non-Hermitian in  ${\cal K}$
but this Hilbert space is
unphysical \cite{SIGMA}.
A precise and detailed mathematical
formulation of this conclusion was given
in survey~\cite{[a2]}:

\begin{lemma} \cite{[a2]} \label{lemma1}
The observable predictions concerning
a closed quantum system
and using
the space-Hamiltonian doublets
$({\cal H},H)$ or $({\cal L},\mathfrak{h})$
are equivalent.
\end{lemma}

 \noindent
In the theories of closed quantum systems
based on equivalences
$({\cal H},H)\ \equiv \ ({\cal L},\mathfrak{h})$
the terminology is still unsettled.
The underlying innovative
formulation of quantum mechanics is often called
quasi-Hermitian \cite{Geyer} or,
more or less equivalently,
${\cal PT}-$symmetric \cite{Carl}
{\it alias\,}
three-Hilbert-space (3HS, \cite{SIGMA})
{\it alias\,}
crypto-Hermitian \cite{Smilga}
{\it alias\,}
pseudo-Hermitian \cite{ali}.
Mathematicians, in contrast, seem to speak just about
non-self-adjoint operators without adjectives \cite{book}.
In what follows,
the acronym 3HS will be mostly employed.

\subsection{Non-uniqueness of Hilbert space ${\cal H}$}

%\newpage
%
%
%
%\subsection{M}

In practice people usually make use of just one
of three alternative
consequences of the equivalence in Lemma \ref{lemma1}.
The first one is trivial: whenever
the evaluation of predictions remains feasible
{\em before\,} preconditioning (\ref{precoxa})
(i.e., in the initial
representation $({\cal L},\mathfrak{h})$),
there is no reason to leave the comfortable
constructive single-space textbook
recipes. The selection of one of the other two options,
i.e., the description of reality
in $({\cal H},H)-$representation
is usually motivated either by the
prohibitively complicated aspects of
Hamiltonian $\mathfrak{h}$, or by a technical
unfriendliness of Hilbert
space ${\cal L}$.

The first version of the ``non-Hermitian'',
$({\cal H},H)-$based option
was made popular, many years ago, by Dyson \cite{Dyson}.
Its key technical ingredient may be seen in
accessibility of an initial
knowledge
of mapping $\Omega$.
Still, even with this knowledge, the not too elementary transition
from the traditional physical Hilbert space ${\cal L}$
to its {\it ad hoc\,} auxiliary partner ${\cal K}$
(in which the Hamiltonian becomes non-Hermitian)
must really very strongly be motivated by
the associated emergent technical simplifications \cite{Geyer}.

In our present paper
we intend to prefer
the second ``non-Hermitian'' option which,
in essence, means
that one wants to work in a maximally user-friendly
Hilbert space ${\cal K}$ from the very beginning.
The analysis then
starts from a presumably
physical diagonalizable
Hamiltonian candidate
$H$ and from its diagonalization
leading to formulae (\ref{bio0}).

It is necessary to add
that when we decided to
set, in our preceding considerations,
$|\psi(t)\kkt=\Theta\,|\psi(t)\kt$,
we simplified our analysis
only at an expense of ignoring the fact that
during the
construction of the basis we
did not take into considerations that either
the
biorthonormality and  completeness relations are both
invariant under
a simultaneous rescaling
of $|n\kt \to |n\kt / \varrho_n$
and $|n\kkt \to |n\kkt \times \varrho_n$.
As we already emphasized in \cite{SIGMAdva}, this
is a well known fact \cite{Geyer} which
implies that
the reconstruction of the physical
Hilbert space ${\cal H}$
(i.e., the formulation of the physical contents of the theory)
is in fact ambiguous and, whenever one wishes to suppress this ambiguity,
the recipe requires more information.

The ``missing'' information may find its origin in physics
(one can introduce some other
observables \cite{arabky},
cf. also reviews \cite{Geyer,Carl}).
Some additional mathematical requirements may also help
(in \cite{lotor}, for example, we proposed to minimize the
anisotropy of the geometry in the physical Hilbert space).
Still, in both contexts the
essence of the problem is that
the innocent-looking rescaling
redefines the metric. Thus,
just one of its special cases was used in Eq.~(\ref{innpro}).
The
general
Hamiltonian-compatible form of the physical metric is
in fact
non-unique and multiparametric,
 \be
 \Theta= \Theta(\vec{\varrho})
 =\sum_n\,|n\kkt\,\varrho_n^2\,\bbr n|\,.
 \label{bio1}
 \ee
This implies  that
also the inner product $(\bullet,\bullet)$ and space ${\cal H}$ become
variable, ambiguous and
$\vec{\varrho}-$dependent.

For practical purposes, there exist multiple methods of removal of
such an unpleasant
non-uniqueness of the theory. Their samples
may be found, e.g., in \cite{Geyer,Carl} or in \cite{lotor}.

\section{The concept of exceptional points\label{section3}}

One of the main distinguishing features of the parameter-dependent
non-Hermitian Hamiltonians $H(\lambda)$ is that
even in the absence of any symmetry (including, in
a very prominent place,
supersymmetry) one can still encounter a degeneracy of
the energy levels
in their spectra \cite{Berry}.
According to Kato \cite{Kato},
such a degeneracy proves to be of a fundamental importance
in perturbation theory. For this reason he conjectured
to call the underlying values of parameters exceptional points
(EPs). In a way depending on the mathematical context
these values $\lambda^{(EP)}$ were characterized either
by their analytic-function background (i.e.,
by their connection with the branch point singularities,
see the next paragraph), or
by the role played in spectral theory.
The latter area of mathematical applications
also led to the Kato's best known definition
of the EPs as the points at which the algebraic
multiplicity of an energy eigenvalue becomes different
from its geometric
multiplicity (see p. 62 in {\it loc. cit.}).

For the practical users of the idea
it is sufficient to know that
at the instant of the EP degeneracy of the energies one
also observes a
parallelization of some of the
eigenvectors.
Still, in the context of physics it took time before
people imagined that
the originally not too emphasized
theoretical as well as experimental
accessibility of
these degeneracies could
open a broad new area of many new EP-related phenomena
\cite{Stellenbosch,Stellenboschb,passage}.

\subsection{The birth of the concept in perturbation theory}

From the strictly historical point of view
the current interest in exceptional points
grew from several independent sources. All of them are,
directly or indirectly,
related to the abstract mathematical concept of an
analytic function $\mathbb{F}(\lambda)$.
One should remember that $\mathbb{F}(\lambda)$
is, in general, multivalued
at its
(complex) variable $\lambda$
so that, strictly speaking, it is not a function
but rather a collection of several single-valued functions
which are
represented, up to a suitable radius of convergence $\lambda_{\max}$,
by the Taylor series. For a quick clarification of this concept
it is sufficient to recall the square-root function  $\mathbb{S}(\lambda)$
which is, in the complex plane of $\lambda$, two-valued.
Its two branches $\mathbb{S}_{\pm}(\lambda)=\pm \sqrt{\lambda}$
are already single-valued functions which coincide
at $\lambda=0$. They may be defined, locally,
by their respective Taylor series ${T}_{\pm}(\lambda)$.
Thus, e.g., at a small complex $\xi$
in $\lambda=\lambda(\xi)=1/2+\xi$
these two
Taylor series are infinite,
 $$
 \pm \sqrt{2}\,{T}_{\pm}(\lambda)=
 1+\xi-{\frac {1}{2}}  \xi  ^{2
}+{\frac {1}{2}}  \xi  ^{3}-{\frac {5}{8}}
  \xi  ^{4}+\ldots \,,
 $$
and they both have the radius of convergence equal to $\xi_{\max}=1/2$.

In Kato's book \cite{Kato} the Taylor series represent
the roots $z_n=z_n(\lambda)$ of
secular equations $\mathbb{P}(z)=\det [H(\lambda)-z\, I]=0$
where
the operator $H(\lambda)$ (or, in simpler cases,
an $N$ by $N$ matrix $H^{(N)}(\lambda)$)
plays the role of quantum Hamiltonian.
The explicit Taylor-series constructions
of its bound-state-energy roots $z_n(\lambda)=E_n(\lambda)=E_n(0)
+\lambda\,E_n'(0)+\ldots$ are then deduced and shown to be
of paramount importance
in practice.
What is essential is the knowledge of
the radius of convergence $\lambda_{\max}$.
A key to the answer is that
the collection of
roots $E_n(\lambda)$ often admits
a unifying analytic-function reinterpretation
$\mathbb{E}(\lambda)$. Its separate
single-valued sheets represent
energy levels $E_n(\lambda)$.
The radius of convergence
$\lambda_{\max}$ acquires an elementary meaning of the distance
of the origin from the nearest EP singularity.

\subsection{EPs in two by two effective Hamiltonians}

In recent review paper \cite{DDNT} the authors emphasized that
the study of certain complicated integrable
models (IM) in 1+1 dimensions
may often profit from the so called
ODE/IM correspondence. The acronym ODE abbreviates
here the
ordinary differential equations, and the benefits
are not only bidirectional
but also deeply relevant in the present context.
This was demonstrated, earlier,
in \cite{DDT} where the
same correspondence formed a background of the
famous proof of reality
of the bound-state energy spectra
of certain ordinary differential
three-parametric quantum Hamiltonians $H$
which were found non-Hermitian in ${\cal K}=L^2({\cal S})$ (here
the symbol ${\cal S}$ stands for a suitable
complex contour).

In a brief addendum \cite{DDTsusy} the authors of the proof
paid more attention to the
structure of the EP boundaries $\partial {\cal D}$ of the
domain of parameters at which the spectrum remains real.
They pointed out that in a small EP vicinity
the passage of parameters through
the boundary can be visualized via an effective two-by-two Hamiltonian
 \be
 H^{(2)}(\delta)=
 \left (
 \begin{array}{cc}
 0&1\\
 \delta&0
 \ea
 \right )\,,
 \ \ \ \ \
  H^{(2)}(\delta)\,
  \left (\begin{array}{c}
 1\\
 \eta
 \ea
 \right )=\eta\, \left (\begin{array}{c}
 1\\
 \eta
 \ea
 \right )\,,
 \ \ \ \ \eta=\eta_\pm=\pm \sqrt{\delta}\,.
 \label{ddtham13}
 \ee
This was finally the simplification which helped them
to clarify some aspects of the reality proof in \cite{DDT}
including, in particular,
the fact that in their model one can also detect
a certain form of hidden
supersymmetry.

All of these observations contributed
to the motivation of our present study.
First of all we imagined that a replacement of
a complicated Hamiltonian
by its ``effective'' two-by-two-matrix simulation
offers a useful and universal mathematical tool of
a qualitative description
of the spectrum near its EP singularity.
Secondly,
the efficiency of the use of the
effective $N$ by $N$ Hamiltonians
seems to survive
even a transition
to the more complicated EP singularities,
i.e., far beyond the class of their comparatively elementary
ODE realizations \cite{BeWu}.
Thirdly,
even if one stays just with the most conventional ODE models,
the effective-matrix simplifications may be expected to
open a viable way towards a
semi-explicit construction
of the physical Hilbert space ${\cal H}$,
i.e., of the last -- and often omitted -- step of the theory.

In the special case of
ordinary differential Schr\"{o}dinger equations,
virtually all of the other available
methods of such a construction of  ${\cal H}$
and $\Theta$
seem to be
prohibitively difficult \cite{117}.
In the approximation by
the two by two matrix simulation (\ref{ddtham13}),
the latter last step remains elementary.
Indeed, once we insert
matrix $H^{(2)}(\delta)$ of Eq.~(\ref{ddtham13})
together with a real and symmetric general
matrix ansatz for the corresponding
Hilbert-space metric $\Theta^{(2)}$ in the hidden Hermiticity condition
(\ref{dieux}) we obtain,
with obvious proof, an exhaustive answer
at any positive $\delta=\eta^2$ and positive $\eta$,
 \be
 \Theta=\Theta^{(2)}(\delta,b)=
 \left (
 \begin{array}{cc}
 \eta&b\\
 b&1/\eta
 \ea
 \right )\,.
 \label{me2}
 \ee

\begin{lemma}
\label{lemma2}
The necessary positivity of metric (\ref{me2})
is guaranteed whenever
its free real parameter $b$ is such that $b^2<1$.
\end{lemma}

%\newpage

\subsection{EPs in crypto-Hermitian models with local
potentials\label{section3b}}

In the physics-oriented literature, by far the most popular class of
illustrative
``non-Hermitian''
Hamiltonians $H$ with real spectra is being obtained via
a judicious modification of some of the standard self-adjoint models
of textbooks.

\subsubsection{Regularized harmonic oscillator\label{rehoho}}

In the above-cited papers \cite{DDNT,DDT,DDTsusy} the authors
studied a specific three-parametric
family of non-Hermitian one-dimensional Schr\"{o}dinger ODEs
describing bound states in a local potential and having the
standard eigenvalue-problem form
$H(M,A,\ell)\,\psi=E\,\psi$
where
 $$
 H(M,A,\ell)=
 -\,\frac{d^2}{dx^2}
+ \frac{\ell(\ell+1)}{x^2} - (ix)^{2M} -A\,(ix)^{M-1}\,
 $$
and where $x$ moved along a suitable {\it ad hoc\,}
complex contour.
Using
sophisticated mathematical methods these authors
contributed to the clarification of the not quite expected fact that
inside a suitable domain of parameter
${\cal D}$ (with, as we mentioned above,
EP boundary  $\partial {\cal D}$)
the bound-state spectrum $\{E_n\}$ proved real and discrete
even though the potential itself was not real.

These observations were preceded by our letter \cite{ptho}
in which we showed that
in the one-parametric special case of $H(1,0,\ell)$
(which appeared to be exactly solvable),
many results become obtainable
via entirely elementary methods.
Also, the exact diagonalizability of the special case with
Hamiltonian $H(1,0,\ell)$ or, equivalently,
the exact solvability of the following Schr\"{o}dinger
ODE bound state
problem with $\alpha=\ell+1/2 \geq 0$,
 \be
 \left (-\,\frac{d^2}{dx^2} + x^2 -2ic\,x
+ \frac{\alpha^2-1/4}{(x-ic)^2} \right )
 \, \varphi(x) =
(E+c^2)  \, \varphi(x), \ \ \ \ \ \varphi(x) \in {\cal K} =
L_2(-\infty,\infty)
 \label{SEho}
 \ee
will prove useful, in what follows,
for illustration of several nontrivial SUSY-EP correspondences.

In a preparatory step let us now briefly summarize some of
the most relevant features
of the bound-state solutions of Eq.~(\ref{SEho}).
Firstly, in terms of the
confluent hypergeometric special functions their
general form
is known,
 \ben
\varphi(x) =  C_+\,(x-ic)^{-\alpha+1/2}e^{-(x-ic)^2/2}\ _1F_1
\left (  (2-2\alpha-E)/4, 1-\alpha; (x-ic)^2
 \right )
+
 \een
 \ben
+
 C_-\,(x-ic)^{\alpha+1/2}e^{-(x-ic)^2/2}\ _1F_1 \left (
 (2+2\alpha-E)/4, 1+\alpha; (x-ic)^2
 \right ).
 \een
Secondly, these functions belong to ${\cal K} =
L_2(-\infty,\infty)$ if and only if
the infinite series terminates. This yields the compact formula
for the wave functions
  \be
\varphi(x) = const. \,(x-ic)^{Q \alpha+1/2}e^{-(x-ic)^2/2} \
L^{(Q \alpha)}_n \left [
 (x-ic)^2
 \right ]\,.
\label{waves}
 \ee
They are numbered by the so called quasi-parity
quantum number $Q=\pm 1$ and by another,
common
quantum number $ n=0,1,\ldots$. The power and exponential factors
are accompanied here by the classical
Laguerre
polynomials,
 \ben
 \ba
 L^{\beta}_0(z)=1,\\
 L^{\beta}_1(z)=\beta+1-z,\\
 L^{\beta}_2(z)=(\beta+2-z)^2 -(\beta+2),\\
 L^{\beta}_3(z)=(\beta+3-z)^3 -3(\beta+3)(\beta+3-z)
 +2(\beta+3),\\
 \ldots \,.
  \ea
 \een
What is most important (cf. \cite{srni})
is that the operator ${\cal Q}$
with quasi-parity eigenvalues $Q$ has later been
reinterpreted as the operator of
charge ${\cal C}$ which defines,
together with conventional parity ${\cal P}$,
the most popular
Hilbert-space special metric
$\Theta = {\cal CP}$ \cite{ali,Carl}.
As a consequence,
the values of energies are all
real whenever $\alpha \geq 0$,
 \be
 E=E_{Q, n}=2 - 2 Q \alpha+4n\,,
 \ \ \ \ \ \ Q = \pm 1\,,
 \ \ \ \ \  n = 0, 1,
2, \ldots
 \,.
 \label{strima}
 \ee
Nevertheless, a warning must follow:
The validity of
the strictly mathematical result (\ref{strima})
of Ref.~\cite{ptho}
{\em does not
imply\,} that all of these real numbers
can really be treated as
bound state energies.

%********** Figure 1 zde
\begin{figure}[h]                     %instead of \begin{figure}[t]
\begin{center}                         %instead of \begin{center}
\epsfig{file=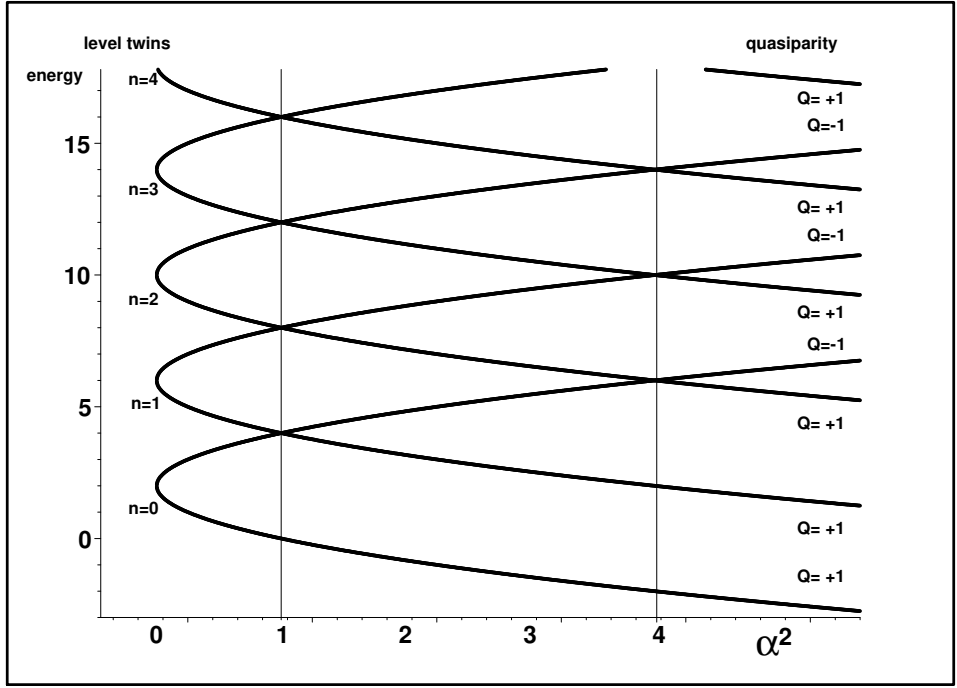,angle=270,width=0.72\textwidth}
\end{center}                         %instead of \end{center}
\vspace{-2mm} \caption{The $\alpha-$dependence of the spectrum
of the singular one-dimensional
harmonic oscillator regularized by a complex shift of
the axis of coordinates.
The conventional
limit with equidistant spectrum
is obtained in the non-singular special case
with $\alpha^2=1/4$. All integers $\alpha$
are exceptional points, $\{\alpha^{(EP)}\}= \mathbb{Z}$.
 \label{reone}}
\end{figure}

The reasons are rather subtle, made explicit only
long after the publication of {\it loc. cit}. Their essence
lies in the emergence of EPs and in the ambiguity of metrics.
The effect is most easily clarified via Fig.~\ref{reone}.
In this picture the leftmost, most obvious EP degeneracy is
easily seen to occur at
$\alpha=\alpha_0^{(EP)}=0$. This is an instant of
complexification,
very often called
spontaneous breakdown of
${\cal PT}-$symmetry \cite{Carl}.
This is a very
special EP mathematical singularity.
In~\cite{denis} it has been given
an alternative physical meaning and name of
quantum phase transition of the first kind.
Schematically, the underlying mechanism
of the complexification is well explained by the
two by two matrix model
(\ref{ddtham13}) of Ref.~\cite{DDTsusy}.
In the limit $\delta \to 0$ this matrix
degenerates to the manifestly non-diagonalizable
(i.e., unobservable and phenomenologically unacceptable)
Jordan-block matrix $J^{(2)}(0)$.

It is less easily seen
that all of the subsequent integers
$\alpha=\alpha_k^{(EP)}=k=1,2,\ldots$
are also exceptional points
(cf. \cite{ptho} for the proof).
In \cite{denis6}
we
explained their specific nature, and we
proposed to call these
unavoided level crossings
without complexification
the quantum phase transitions
of the second kind.
In the light of the
Laguerre-polynomial solvability of the present model
the quantum phase transition degeneracies
of the second kind can even be given
the form of exact identities
 $$
  L^{(-1)}_{n+1}\left [ (x-ic)^2 \right ]
  =-(x-ic)^2\,L^{(1)}_n\left [ (x-ic)^2 \right ]
  $$
etc.

%\newpage

\subsubsection{Regularized three-body Calogero model}

In \cite{Tater} we re-interpreted Schr\"{o}dinger Eq.~(\ref{SEho})
as a two-body special case of an integrable $A-$body Calogero model
of the rational ${\cal PT}-$symmetric $A_n$ series
\cite{Calogero,Fring}.
Although the discussion in {\it loc. cit.}
did not involve the question of EPs,
we are now re-attracting attention to this
family of models because their
exact solvability
again opens the way
of our understanding
the presence, localization and properties of the EPs
in a less elementary setting.

For the sake of brevity let us only recall here the
first nontrivial
three-body version of the Hamiltonian
 \ben
 \tilde{H}^{(3)}
 = \sum_{i=1}^{3}\ \left [-
\frac{\p^2}{\p{x_i}^2}
 + \frac{3}{8}\,\omega^2
\,x_i^2  \right ]
 +\frac{g}{ (x_1-x_2)^{2}}
 +\frac{g}{ (x_2-x_3)^{2}}
 +\frac{g}{ (x_3-x_1)^{2}}\,.
  \een
Interested readers may
find all of the technical details and formulae in \cite{Tater}.
Here we will only recall a few most relevant features of the model
and add a few EP-related remarks.

First of all we will
fix the units and set
$\omega^2=\frac{8}{3}$ yielding
the compact formula
 \be
 E^{(\pm)}_{n,k}=4n+6k\pm 6\alpha + 5
 ,\ \ \ \  \ \ \ \alpha=\frac{1}{2}\sqrt{1+2g}>0, \ \ \ \ \ n,k=0,
 1, \ldots\
 \label{fifor}
 \ee
which defines the spectrum.
From our present
point of view such a formula demonstrates, immediately,
that the EP singularites may be expected to occur,
in these integrable models,
at any number of particles $A$. At the same time,
an increase in the complexity of the spectrum at $A=3$
(made explicit by the occurrence of an additional
quantum number $k$) indicates that not all of the
unavoided level crossings must necessarily be the EP
singularities. Some of them just reflect
the presence of a conventional (e.g.,
rotational \cite{Tater}) symmetry in the
corresponding partial differential
Schr\"{o}dinger equation.
A conversion of formula
(\ref{fifor}) into a picture (sampled by Fig.~\ref{rethree} here)
just reconfirms that in contrast to the
model of the preceding subsection, some of
the
energy levels
become degenerate
due to the classical symmetries of interaction
rather than due to the
emergence of an EP singularity.

%********** Figure 2 zde
\begin{figure}[h]                     %instead of \begin{figure}[t]
\begin{center}                         %instead of \begin{center}
\epsfig{file=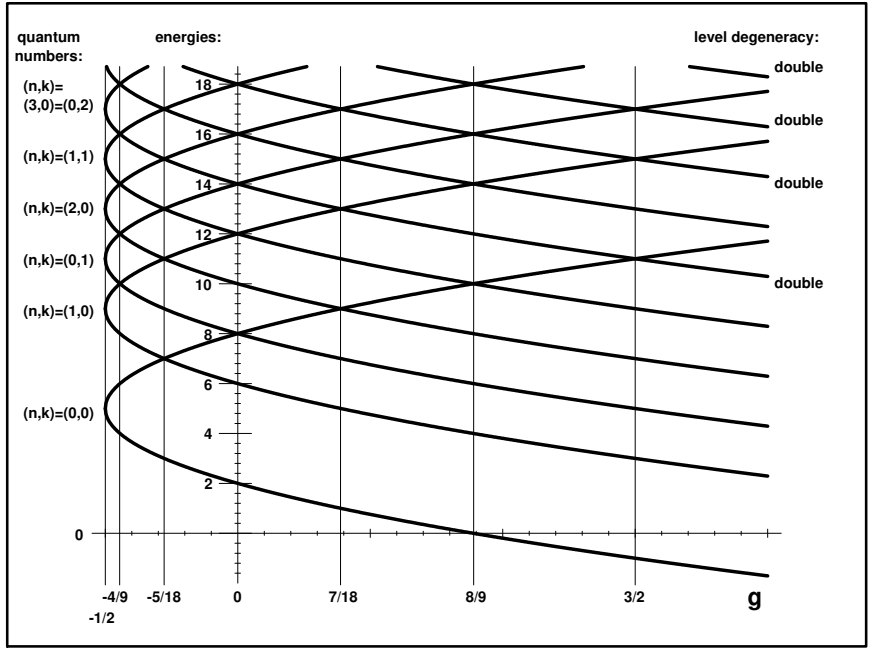,angle=270,width=0.72\textwidth}
\end{center}                         %instead of \end{center}
\vspace{-2mm} \caption{The degeneracies and
coupling-dependence of the spectrum
of the non-Hermitian three-particle Calogero
model of Ref.~\cite{Tater}. Levels
$E_{(n,k)}= 5+4n+6k \pm 3\,\sqrt{1+2g}$
are numbered by a pair of integers $n,k=0,1,\ldots$.
All of the visible
exceptional points $g^{(EP)}$ are marked by vertical lines.
 \label{rethree}}
\end{figure}
%($E_{(n,k)}= 5+4n+6k \pm 3\,\sqrt{1+2g}$)

At the higher quantum numbers
this phenomenon really does occur
in a way mentioned, in Fig.~\ref{rethree},
in the right column of comments.
Unfortunately, we do not have enough space here
for a sufficiently exhaustive discussion of the
differences or analogies
between the
qualitatively slightly different dynamical scenarios with
$A=2$ and $A=3$ or $A\geq 3$.
In fact, one of our main reasons
is that after one turns attention to the supersymmetric
extensions of these models, the number of open questions
starts growing rather quickly
even during the study of the
perceivably simpler illustrative calogerian model with $A=2$.

%\newpage

\section{The concept of hiddenly Hermitian
supersymmetry\label{section4}}

Even though the SUSY-based arrangement of
existing elementary particles
into supermultiplets \cite{Likh}
did not work in particle physics
\cite{Witten},
the idea itself proved extremely successful after
its transfer to
quantum mechanics.
It helped
to
classify a family of exactly solvable models
\cite{Khare,Bijansusy}.
In the words of the latter reference,
``in a supersymmetric
theory commutators as well as anticommutators appear in the
algebra of symmetry generators''.
For our present purposes we shall use
the explicit forms of these generators
forming the specific Lie superalgebra sl(1$|$1)
(see Eq.~(\ref{grlieal}) below).

The same or similar ideas were also actively used during the birth of
quasi-Hermitian theories \cite{DDTsusy,Bendsusy,Ioffe,Cannata}.
Surprisingly enough, during or after
all of these promising developments,
attention remained restricted
to the spectral problem,
i.e., to the solution of
the non-Hermitian Schr\"{o}dinger equations in ${\cal K}$.
Virtually no results were obtained in the direction of
construction
of the corresponding physical Hilbert space
of states ${\cal H}$, i.e., of the amended
operator $\Theta \neq I$ of
the correct physical Hilbert-space metric.

The latter gap in the literature
and, in particular, in the SUSY-related implementations
of the 3HS theory
is in fact not too surprising because
the task is known to be truly difficult \cite{ali,117}.
At the same time, the knowledge of the geometry of
${\cal H}$ becomes crucial whenever the metric $\Theta$
deviates from the unit operator ``too much''
(see, e.g., the thorough analyses of this topic in
\cite{Viola,admissible}). For this reason,
at least the use of
qualitative considerations and approximate methods is needed
when one decides to turn attention to the dynamical regime
close to an EP extreme.

\subsection{Harmonic oscillator example\label{HOE}}

Let us now return to our quasi-Hermitian harmonic-oscillator model
(\ref{SEho}) and let us amend slightly the notation:
In terms of the complex coordinate
 \be
 r = r(x)=x-i\,c, \ \ \ \ \ \ x \in (-\infty,\infty)\,,
 \ \ \ \ c >0
 \label{regularization}
 \ee
we shall consider the
regularized quasi-Hermitian harmonic-oscillator
Hamiltonians
 \be
  H^{(\xi)}= -
 \frac{d^2}{dr^2}
 +
 \frac{\xi^2-1/4}{r^2}
 + r^2
 \label{SEha}
 \ee
and their two shifted versions
 \be
 {H}_{(L)} = {H}_{}^{(\alpha)} -2\gamma-2,
 \ \ \ \ \ \
 {H}_{(R)} = {H}_{}^{(\beta)} -2\gamma
  \label{Mtt}
 \ee
such that ${\alpha}=|\gamma|$ and $\beta=|\gamma+1|$.
In the light of the results of paragraph \ref{rehoho}
we may conclude that all of the integer values of $\gamma$
are exceptional points, $\{\gamma^{(EP)}\}=\mathbb{Z}$.
These values leave
both of our sub-Hamiltonians
non-diagonalizable so that they have to be excluded from our
consideration as manifestly unphysical.

Under this constraint, in terms of the so called superpotential
 \be
 W=W_{}^{(\gamma)}(r) =
 r-\frac{\gamma+1/2}{r}\,
 \label{N}
 \ee
we may define operators
 $A=\p_x+W$ and $B=-\p_x+W$
and form the set of generators
 \ben
 {\cal G}= \left [ \begin{array}{cc} H_{(L)}&0\\ 0&H_{(R)}
 \ea
 \right ]
, \ \ \ \ \ \
 {
 \cal Q}=\left [
 \begin{array}{cc} 0&0\\ A^{}&0
 \ea
 \right ],
 \ \ \ \ \ \
\tilde{\cal Q}=\left [
 \begin{array}{cc}
0& B^{}
\\
0&0 \ea \right ]\
 \een
of the well known Lie superalgebra sl(1$|$1),
 \be
 \{ {\cal Q},\tilde{\cal Q}
\}={\cal G} , \ \ \ \ \ \ \{ {\cal Q},{\cal Q} \}= \{ \tilde{\cal
Q},\tilde{\cal Q} \}=0, \ \ \ \ \ \ \ \ [ {\cal G},{\cal Q} ]=[
{\cal G},\tilde{\cal Q} ]=0.
 \label{grlieal}
  \ee
Interested readers may find an exhaustive description
of details in \cite{2002}
[cf. also the present Eq.~(\ref{strima}) for energies].

In this setup our present intention is to pay
attention to the
occurrence and role of the EPs which were not
properly treated
in paper \cite{2002}. We will fill the gap
in what follows.
For such a purpose we
will only need to recall,
in explicit form, the
existing results concerning the spectrum
of the $\gamma-$dependent super-Hamiltonian
${\cal G}={\cal G}(\gamma)$.
Its parameter $\gamma \in \mathbb{R}$
taken from superpotential (\ref{N}) is in fact the key quantity
which determines the behavior of the (super)energies.
In the light of this fact let us now split the real line of
$\gamma$s into three subintervals.

\subsubsection{Central interval of $\gamma \in (-1,0)$ with
$\alpha=-\gamma$ and $\beta=\gamma+1$.}

As long as our super-Hamiltonian
${\cal G}(\gamma)$ is a direct sum of its ``left'' and ``right''
sub-Hamiltonians $H_{(L)}$ and $H_{(R)}$,
we may denote the real $N-$th-level energy values
(with $N=0,1,\ldots$) by the respective pairs of
symbols $[E^{(L)}_N,E^{(R)}_N]$ (at
the positive quasi-parity $Q=+1$)
and $[F^{(L)}_N,F^{(R)}_N]$ (at
the negative quasi-parity $Q=-1$).
In such an arrangement we may conclude:

\begin{lemma}
\label{lemma3}
For $\gamma=\xi-1/2 \in (-1,0)$ the ground state energy
$E^{(L)}_0=0$ is non-degenerate while the rest of the
spectrum is doubly degenerate and such that
 $$
 E^{(R)}_N=F^{(L)}_N=4N+2-4\xi <
 F^{(R)}_{N}=E^{(L)}_{N+1}=4N\,,\ \ \ \ N=0,1,\ldots\,.
 $$
\end{lemma}
 \noindent
Marginally, it is possible to add that the spectrum becomes
equidistant at $\gamma +1/2=\xi=0$.

\subsubsection{The right half-line of
$\gamma =\alpha \in (0,\infty)$ with
 $\beta=\gamma+1$.}

As we indicated above, all of the integer values of $\gamma$
have to be excluded from our considerations because
at these values
we have $\gamma=\gamma^{(EP)} \in \mathbb{Z}$
so that our quantum model loses any acceptable
probabilistic interpretation \cite{ali}.

\begin{lemma}
\label{lemma4}
In every admissible open subinterval of $\gamma=\alpha=\beta-1
\in (K,K+1)$ with integer $K\geq 0$
the spectrum contains a single non-degenerate
excited-state energy $F^{(L)}_0=0$
which separates the energy spectrum
into the negative ordered degenerate doublets
 $$
  E^{(L)}_N= E^{(R)}_N=-4\gamma+4N\,,
  \ \ \ N=0,1,\ldots,K
 $$
and the ordered
pairs of the positive degenerate doublets
 $$
  E^{(L)}_N= E^{(R)}_N=4N-4\gamma<
    F^{(R)}_{N-1}= F^{(L)}_N=4N\,,
  \ \ \ N=K+1,K+2,\ldots\,.
 $$
\end{lemma}
 \noindent
A note can be added that
the ground state is now doubly degenerate.
At the half-integer values of $\gamma = K+1/2$ the
positive
doubly degenerate part of the spectrum becomes
equidistant again.

\subsubsection{The left half-line of $\gamma \in (-\infty,-1)$ with
$\alpha=-\gamma$ and $\beta=-\gamma-1$.}

\begin{lemma}
\label{lemma5}
In every admissible open subinterval of $\gamma=-\alpha=-\beta-1
\in (-K-1,-K)$ with integer $K\geq 1$
the spectrum contains a single non-degenerate
ground-state energy $E^{(L)}_0=0$
followed, at $K \geq 2$,
by an anomalous $(K-1)-$plet
of degenerate pairs of the lowest excited-state
energies
 $$
  E^{(L)}_{N+1}= E^{(R)}_N=4N+4\,,
   \ \ \ N=0,1,\ldots,K-2.
 $$
The rest of the spectrum is then formed,
at any $K \geq 1$,
by the ordered
pairs of degenerate doublets
 $$
  E^{(L)}_{N+1}= E^{(R)}_N=4N+4 <
    F^{(L)}_{N}= F^{(R)}_N=4N+4+4\beta\,,
  \ \ \ N=K-1,K,\ldots\,.
 $$
\end{lemma}

 \noindent
At the half-integer values of $\gamma = -K-1/2$ the
equidistance now characterizes the non-anomalous
part of the doubly degenerate spectrum,
i.e., the spectrum without the lowermost
$(2K-1)-$plet of levels (counting their
SUSY-related degeneracy).

\begin{cor}
In any preceding crypto-Hermitian harmonic-oscillator
realization of SUSY algebra (\ref{grlieal})
the range of the admissible parameters $\gamma$ always remains restricted to
a finite interval of unit length,
with the EP boundaries given by
the two neighboring integers.
\end{cor}

%\newpage

\subsection{Regularized Hermitian limit}

In 1984 Jevicki and Rodrigues \cite{JR} pointed out that
formalism of supersymmetric quantum mechanics breaks down
for many phenomenologically interesting potentials with
singularities (i.e., e.g., with
the repulsive centrifugal-like barriers).
In \cite{2002} we showed that
what could help is
a complex-shift regularization, and
we asked what happens with the SUSY structures in
the limit $c \to 0^+$.
Our considerations were illustrated by the
regularized harmonic oscillator of Eq.~(\ref{SEho}).
Thanks to the exact solvability of this model
we were able to summarize our observations in
Table Nr. 3 of Ref.~\cite{2002} and in the appended
spectrum-explaining comments.

From the present point of view
the latter results
deserve a comment. The point is that
the centrifugal singularity remains regularized
at any non-vanishing shift-parameter $c>0$.
In {\it loc. cit.} we were interested also in the behavior of the
system in the limit  $c \to 0^+$.
In this limit the regularization is removed
so that the norm of at least some of the $c>0$
solutions diverges.
In {\it loc. cit.} we
reviewed the situation and we
managed to show that
some of the solutions still remain acceptable by obeying
the emergent standard boundary conditions in the origin.
The resulting, drastically restricted family of
normalizable $c=0$ bound states
is displayed in Fig.~\ref{retwo}.

%********** Figure 3 zde
\begin{figure}[h]                     %instead of \begin{figure}[t]
\begin{center}                         %instead of \begin{center}
\epsfig{file=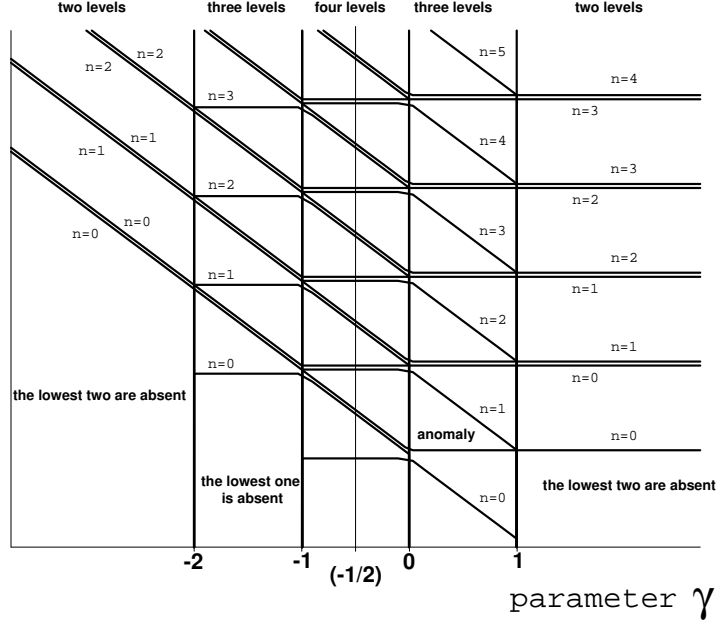,angle=270,width=0.72\textwidth}
\end{center}                         %instead of \end{center}
\vspace{-2mm} \caption{Degeneracies
of the $c\to 0$ spectra
of the quasi-Hermitian and
supersymmetric  harmonic
oscillator (\ref{SEho}). In the conventional
regular $\gamma=-1/2$ special case
the supersymmetric Hamiltonian becomes self-adjoint.
Only four exceptional points
$\gamma \in \{-2,-1,0,1\}$
remain phenomenologically relevant
since
the sets of bound states become incomplete out of the
innermost admissible interval of
$\gamma \in (-1,0)$.
 \label{retwo}}
\end{figure}

%\newpage

\section{Discussion\label{section5}}

\subsection{SUSY and local potentials}

Whenever the local potentials are being involved
(which seems to be one of the characteristic aspects of
many successful applications of various SUSY models),
it will be absolutely necessary to return
to the
older \cite{Dieudonne}
as well as newer \cite{Viola,Petr}
mathematical criticism of the
very foundations of the 3HS formalism.
Indeed, also in this
more formal direction
of research
many
important questions remain open
at present \cite{ATbook}.

In the field of mathematics,
several interrelated challenges also
survive concerning the constructions of the
Hamiltonian-dependent
Hilbert-space metrics $\Theta=\Theta(H)$ \cite{lotor}.
The abstract freedom of our choice between alternative
formulations of quantum mechanics
(sampled by the pair $({\cal H},H)$
and $({\cal L},\mathfrak{h})$
as mentioned in
Lemma \ref{lemma1})
is of a very restricted practical use.
In applications,
such a choice is usually more or less unique,
determined by
the criteria of feasibility of the
constructions and calculations.

This is a pragmatic attitude which is being accepted
in virtually all of the applied hidden-Hermiticity
settings. A decisive factor leading to the choice of a specific
representation $({\cal H},H)$ is usually found in the
user-friendly nature of the Schr\"{o}dinger equations when solved
in ${\cal K}$.
In this sense the field of SUSY quantum mechanics
seems particularly suitable and promising
fur future reseaarch, especially
due to its deep algebraic as well as analytic background.

\subsection{Analyticity and EPs}

In the conventional quantum physics
using self-adjoint Hamilonians
the EP singularities
are often
perceived as interesting
and, perhaps, mathematically useful
but
not too phenomenologically important
complex-plane points of coincidence
of an $n \neq m$ pair of
the analytically continued bound state energies,
$E_n(\lambda^{(EP)})=E_m(\lambda^{(EP)})$
\cite{BeWu,Alvarez}.
In a historical perspective this encouraged
the study of
operators of observables
(i.e., most often, Hamiltonians) which were
{\em analytic functions\,}
of their parameters, $H(\lambda)=H(0)+\lambda\,H'(0) + \ldots$.
Such an assumption proved to be a natural source of
overlaps between perturbation-theory mathematics and the
PT-symmetry-related physics \cite{BT,Geza}.
Typically, this helped to clarify the connections between
the time-independent
Schr\"{o}dinger equations
 $
 H(\lambda)\psi_n(\lambda)=E_n(\lambda) \psi_n(\lambda)
 $
and their solutions
based on the power-series Rayleigh-Schr\"{o}dinger ansatzs
 $$
 \psi_n(\lambda)=\psi_n(0)+\lambda\,
 \psi'_n(0)+ \ldots
 $$
and
 $$
 E_n(\lambda)=E_n(0)+\lambda\,
 E^{(1)}_n+ \ldots\,.
 $$
In a way discussed by Kato
\cite{Kato},
at least a subset of all of the (real or complex) energy eigenvalues
$E_n(\lambda)$ may be then interpreted
as evaluations of a {\em single\,} analytic,
multisheeted function $\mathbb{E}(\lambda)$.

Kato also emphasized that
a nice
explicit illustration
of the related mathematics may be then provided
by certain {\it ad hoc\,} finite-dimensional
$N$ by $N$ matrix toy-model Hamiltonians $H^{(N)}(\lambda)$.
In the present context of the SUSY - EP
correspondence
this could inspire a further study of
various quasi-Hermitian SUSY
Hamiltonians
and, in particular, their
unphysical EP limits
(plus their small perturbations -- see, e.g.,
a few related remarks in
\cite{passage})
in their canonical
block-diagonal
representations.
For example, in a small vicinity of unperturbed
$\alpha = \alpha^{(EP)}_0=0$
in Eq.~(\ref{SEho}) one could try to replace,
approximately, the
exact
ODE Hamiltonian $H^{(\alpha)}$
of Eq.~(\ref{SEha}) by its
block-diagonal limit
 $$
 H^{(\alpha)}=
 \left(
 \begin{array}{cc|cc|cc}
 2&1&0&0&0&\ldots\\
 0&2&0&0&0&\ldots\\
 \hline
 0&0&6&1&0&\ldots\\
 0&0&0&6&0&\ldots\\
 \hline
 0&0&0&0&10&\ldots\\
 \vdots&\vdots&\vdots&\vdots&\ddots&\ddots
 \ea
 \right ) + {\cal O}(\alpha)
 $$
where the blocks would be just the Jordan-block
matrices
$J^{(2)}(\eta)$
obtained form the matrix of Eq.~(\ref{ddtham13})
in its non-diagonalizable extreme,
$J^{(2)}(\eta)=\lim_{\delta \to 0}H^{(2)}(\delta)
+ \eta\,I^{(2)}$.

In the future applications
involving some less elementary and, in particular,
some supersymmetric
models one may expect to obtain various less schematic
and mathematically more interesting
EP-related matrix structures.
They could serve
as approximations
of realistic Hamiltonians
along the lines studied, e.g., in \cite{admissible}.

\subsection{Outlook}

In our paper we skipped all outlines
of physics behind the 3HS theory,
and also of the multiple early contacts
of this theory with SUSY
(an extensive information on this topic
may be found, e.g., in \cite{MZbook}).
In any forthcoming
phenomenology-oriented analysis
more attention should certainly be paid,
therefore,
to the interlaced SUSY and EP aspects
of the constructions of
the
physical Hilbert spaces of states ${\cal H}$
as well as
to the role played by some other,
non-Hamiltonian
operators of the
hiddenly Hermitian observables.

During the future developments of the field of the
SUSY-EP contacts
one should certainly return, last but not least, also to
the realistic-physics  inspiration provided by
the Dyson-inspired studies
of non-SUSY
interacting boson Hamiltonians $H$
as well as by the recently revealed
Dyson-unrelated
hiddenly Hermitian but non-SUSY
parallel developments
involving, for example,
the clusters-coupling Hamiltonians
$H$ \cite{Bishop}.

%\newpage

\section*{Acknowledgments}

The author acknowledges the financial support from the
Excellence project P\v{r}F UHK 2020.

\end{document}